# Nonlocality is the missing design rule for topological photonics

Fatemeh Davoodi[1,2]
[1]Nanoscale Magnetic Materials, Institute of Materials Science, Kiel University, 24143, Kiel, Germany
[2]Kiel Nano, Surface and Interface Science KiNSIS, Christian Albrechts University, Kiel, Germany
fda@tf.uni-kiel.de

**Abstract**
Topological photonics has provided powerful design rules for routing and localizing light through geometry, symmetry and engineered coupling. However, fabricated nanophotonic and plasmonic devices rarely realize only the local or short-range interaction networks assumed in compact topological models. Long-range near-field coupling, retardation, radiation leakage, substrate-assisted hybridization, material dispersion and fabrication disorder can reshape the optical modes that are interpreted as topological. In this Perspective, I argue that nonlocality should be treated as a design parameter rather than as a residual perturbation. This shift requires moving from ideal phase labels toward calibrated interaction models, finite-structure observables, robustness maps and graded confidence measures. I discuss how full-wave simulations, experiments and physics-informed learning can connect geometric design variables to effective electromagnetic interaction networks. Such calibrated workflows can clarify when a topological design rule is reliable, when it fails, and how nonlocal coupling can be exploited for robust nanophotonic devices.

Topological photonic structures are commonly designed using local or short-range interaction models, although their physical implementations can support substantial long-range electromagnetic coupling. This Perspective identifies interaction nonlocality as a design parameter that should be quantified, calibrated, and incorporated into device validation. Geometry-dependent coupling models, finite-structure observables, robustness maps, and graded confidence measures can connect ideal topological classifications to experimentally implementable nanophotonic systems.

Topological photonics has established a powerful design principle: geometry can control where light propagates, where it localizes, and how strongly it resists selected perturbations [1,2]. A change in coupling, symmetry, lattice arrangement, or synthetic gauge field can produce modes localized at an edge, corner, hinge, surface, defect, or interface [2,3]. This language is powerful because it converts complicated electromagnetic structures into simple design rules. It can also mislead. A topological lattice diagram is not, by itself, evidence that the fabricated optical structure realizes the intended topological Hamiltonian. Nanoparticles, resonators, waveguides, cavities, metasurface elements, and plasmonic building blocks interact through electromagnetic fields whose range, phase, polarization content, and strength depend on wavelength, geometry, material dispersion, substrate environment, radiation channels, and fabrication tolerance [3-7]. Maxwell's equations do not impose the locality assumed by many compact models. This is the hidden nonlocality of topological photonics: the part of the electromagnetic interaction network that is absent from the schematic model but present in the device [3-7]. It includes radiation-mediated coupling, long-range near-field interaction, substrate-assisted hybridization, collective scattering, and geometry-dependent mode mixing [4-7]. These interactions can shift bands, hybridize boundary and bulk modes, alter localization lengths, and broaden what was expected to be a sharp boundary between topological and trivial response [4-6]. The issue is particularly visible in topological photonics because boundary states provide a sensitive test of the realized interaction network. But it is not limited to topological structures. The same nonlocality controls collective resonances in metasurfaces, band formation in photonic crystals, hybridization in plasmonic crystals, and mode formation in coupled-resonator arrays [4,7]. As nanophotonic platforms move from proof-of-principle demonstrations toward functional devices, treating nonlocality as a calibrated design variable rather than as a residual perturbation is becoming essential [7,8]. The next challenge for topological nanophotonics is therefore not only to classify more lattice models, but to determine which interaction network a real optical structure realizes, how that network modifies the intended mode, and whether the claimed topological function survives under realistic electromagnetic coupling.

## A local model is not a device criterion

Local and short-range models remain indispensable. They reveal which symmetries matter, which couplings should be tuned, and why certain modes may become insensitive to selected perturbations [1,2]. Without them, topological photonics would lose much of its conceptual clarity. The problem begins when the model is used as a complete device criterion. A fabricated nanophotonic or plasmonic structure does not only ask whether an ideal model satisfies a clean topological condition. It asks whether the actual electromagnetic object supports the intended optical response after long-range coupling, dissipation, radiative leakage, spectral detuning, finite-size hybridization, and fabrication uncertainty are included. These effects are not secondary corrections to the measured device. They help define the mode that will be observed or used. This distinction is especially important for nonlocal photonic modes. Such modes emerge from strong mutual field interactions between individual resonators. They are not isolated resonator modes repeated periodically, nor are they always captured by adding a small correction to a local model. Their frequency, linewidth, field distribution, radiation pattern, and localization length are determined by the full interaction network [4,6,9,10]. In topological systems, this network controls edge, corner, hinge, surface, or interface localization. In non-topological systems, it can control collective lattice resonances, Fano modes, bound states in the continuum, superradiant and subradiant modes, and plasmonic or dielectric hybrid modes [4,7]. The problem becomes sharper in hybrid systems. Two-dimensional material heterostructures provide a route to material-dependent optical response through layer composition, material dispersion and electromagnetic coupling [6,7,11-15]. These responses add optical degrees of freedom that are not captured by geometry alone and introduce new interaction channels [11,12]. The optical mode is no longer determined only by the patterned photonic geometry, but by a coupled electromagnetic-material network. The relevant design quantity is therefore not only the label assigned by an ideal theory. It is the persistence of the desired electromagnetic response under the interactions that the structure physically supports.

## From model classification to electromagnetic observables

A more reliable design procedure should separate two questions that are often compressed into one. The first is conceptual: what simplified model explains the intended mechanism? The second is operational: what electromagnetic observable demonstrates that the simulated or fabricated structure realizes that mechanism? For topological photonics, the observable may be localization at an edge, corner, hinge, surface, defect, or interface, together with spectral isolation and robustness against disorder [1-3,6]. In chiral or non-reciprocal platforms, it may include propagation direction and suppression of backscattering [1,3,9,10]. In non-Hermitian platforms, it may include linewidth, gain-loss balance, and mode hybridization. The correct diagnostic depends on the physical function of the device, not only on the mathematical label assigned to an ideal model. The same principle applies beyond topology. In metasurfaces, the relevant observable may be wavefront fidelity, diffraction efficiency, polarization conversion, angular tolerance, bandwidth, or suppression of unwanted channels [7]. In photonic crystals, it may be band dispersion, gap formation, slow-light behavior, defect-mode confinement, or radiation leakage. In plasmonic crystals, it may be extinction, scattering, absorption, resonance splitting, linewidth, field enhancement, or spatial energy distribution [4,12]. The point is simple. If the optical function is collective, the validation must also be collective. Reporting only the intended unit-cell response, the nearest coupling, or an ideal band property is not enough.

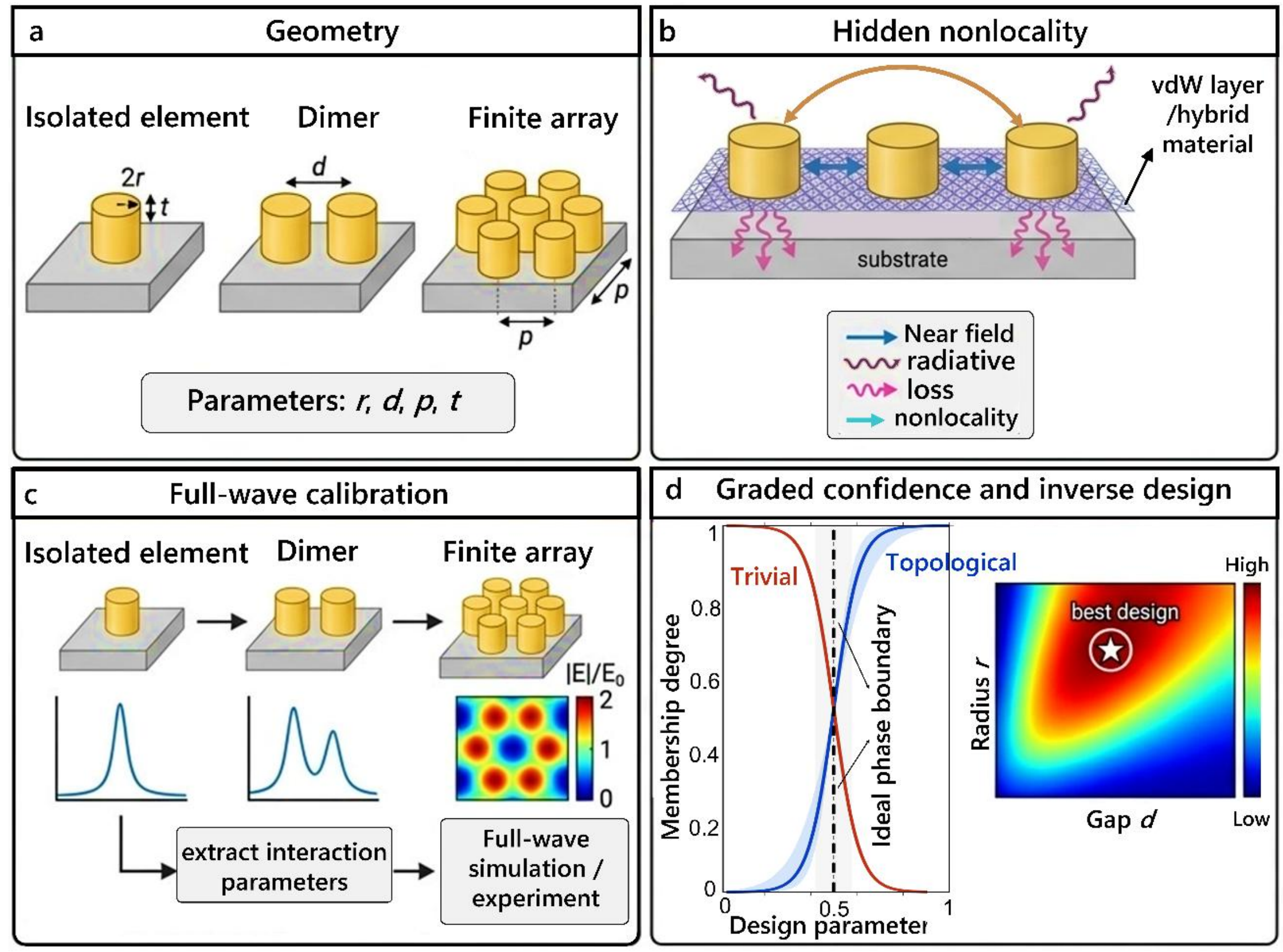

**Figure 1 | From geometric topology to calibrated nonlocal photonic design. a,** A photonic structure is first specified through geometrical variables, including resonator radius *r*, gap *d*, lattice period *p* and thickness *t*. **b**, In realistic nanophotonic or hybrid-material platforms, the effective interaction network is not limited to nearest-neighbor coupling. Near-field coupling, radiative coupling, material-mediated response and loss introduce hidden nonlocal channels. **c**, Full-wave simulation or experiment can calibrate this interaction network by comparing isolated elements, dimers and finite arrays through spectra and field distributions. **d**, The calibrated model replaces a sharp binary phase boundary with graded confidence in trivial, topological and uncertain response classes, enabling robust inverse design over geometric parameters.

## From binary phase labels to graded phase confidence

A further consequence of nonlocality is that the boundary between topological and trivial response need not remain operationally sharp. In an ideal model, a phase boundary can often be drawn as a clean line. In a finite electromagnetic structure, long-range coupling, loss, disorder, spectral overlap, and fabrication uncertainty can broaden this line into a transition region [4-6]. Within this region, the optical response may contain partial boundary character, incomplete spectral isolation, or mixed bulk-boundary hybridization. A binary label can then be less informative than a confidence measure. A design may be strongly topological, strongly trivial, or located in an ambiguous region where the device function depends sensitively on nonlocal coupling, disorder, loss, or calibration uncertainty. Such an ambiguous region should not automatically be dismissed as numerical noise. It may be the operational phase boundary of the actual electromagnetic device. Here, graded phase confidence is introduced as a practical description of this operational transition region. Rather than assigning each geometry only to "topological" or "trivial," one can assign degrees of membership to topological, trivial, and uncertain response classes. This does not replace topological invariants, full-wave simulations, or measurements. It provides a practical way to report where a topological design rule is reliable, where it fails, and where the finite electromagnetic response lies between the two.

## What should be reported

For topological nanophotonic and plasmonic structures governed by collective coupling, four quantities should be reported. A schematic lattice diagram and an ideal phase argument are not sufficient when the physical platform supports longer-range electromagnetic interactions. First, the realized interaction model should be specified. The retained coupling channels should be stated, and neglected channels should be justified. If the design assumes local unit-cell behavior, weak intercell interaction, nearest-neighbor coupling, or negligible radiation-mediated coupling, that approximation should be supported by full-wave simulation, experiment, or a controlled estimate of coupling decay [4,7]. Second, the observable used to validate the design should match the physical function. In a topological structure, this means showing that the relevant mode is localized at the intended boundary, interface, defect, hinge, or corner, and that it remains distinguishable from bulk, radiative, or lossy channels. A topological label alone is not a device measurement. Third, robustness should be mapped rather than asserted. A single optimized geometry is less informative than a landscape showing how the optical response changes under nonlocal coupling, disorder, detuning, loss, wavelength shift, and fabrication uncertainty. A mode may be protected against one perturbation but fragile against another. A robustness map makes that distinction visible [8,16]. Fourth, if the transition between response classes is broadened, the ambiguous region should be reported rather than hidden. A calibrated phase diagram can include phase-confidence maps with continuous membership in topological, trivial, and uncertain classes. This does not weaken the topological claim. It states more precisely where the claim is operationally reliable.

## Learning without losing the physics

Machine learning can accelerate nanophotonic design, but it can also hide the physical variables that matter. A surrogate model may predict a spectrum, classify a phase, or optimize a geometry while providing limited information about which interaction pathway controls the result [16]. This is a weakness when the target response arises from nonlocal coupling. A more suitable approach is physics-native inference. The model should keep the relevant electromagnetic variables explicit: coupling strengths, detuning, radiation channels, boundary weights, modal overlaps, field-energy distributions, and geometry [16]. The outputs should remain interpretable observables rather than detached labels. Quantum equilibrium propagation provides one possible implementation of this principle [17]. In a Hamiltonian-native formulation, a physical coupled-mode or lattice Hamiltonian is connected to a compact trainable sensor sector [17]. The input variables remain physical couplings or geometry-calibrated parameters. The outputs are expectation values associated with chosen observables. Weak nudging supplies the training signal while preserving the separation between physical interactions, trainable sensor couplings, and measured response [17]. The central point is not that the classifier is quantum-inspired. It is that the inference procedure preserves the physics of collective coupling. This structure can be used beyond one specific lattice model. A sensor may distinguish boundary, bulk, surface, or interface sectors in a topological lattice, quantify collective-mode confidence in a resonator array, identify a target hybridized resonance in a plasmonic crystal, or evaluate the stability of a metasurface response when inter-element coupling is non-negligible. In each case, the learned quantity should be an operational optical response. The output of such an approach is not only a prediction. It is a response landscape. Averaging that landscape over coupling distributions, geometry tolerances, or disorder produces a robustness map. If the response classes overlap, the same landscape can be converted into graded phase confidence. This turns nonlocal interaction from an uncontrolled deviation into a design coordinate.

## Geometry closes the loop

Photonic devices are fabricated from geometries, not from abstract coupling constants. A workflow that stops at an effective model remains incomplete unless its parameters are calibrated against electromagnetic simulation or experiment [4,11,12]. For plasmonic and resonant nanophotonic structures, calibration can begin with isolated and paired building blocks. The isolated element defines the local resonance. The pair or cluster reveals the interaction-induced response: resonance shifts, linewidth changes, mode splitting, line-shape deformation, polarization mixing, and near-field redistribution. These signatures can be converted into effective interaction parameters through constrained fitting, full-wave inversion, or physically motivated regression. Once this geometry-to-interaction relation is available, particle size, spacing, lattice period, substrate index, wavelength, and fabrication tolerance can be propagated into the effective model. The same framework can then be used forward, to test whether a proposed geometry supports the desired collective mode, or backward, to search for geometries that make the mode robust. This step is not technical bookkeeping. It is what connects a conceptual design rule to a usable nanophotonic device. Without it, topology remains a label attached to a schematic structure. With it, topology becomes a calibrated statement about an electromagnetic object.

## Toward calibrated topological photonics

Topological photonics should not abandon ideal models. They remain the language in which boundary localization, symmetry, chirality, and robustness are understood. But a model becomes device physics only when it is calibrated to the electromagnetic interactions of the structure being designed. The same lesson applies more widely across nanophotonics. In metasurfaces, local periodic approximations can fail when neighboring elements interact strongly or when the phase profile varies rapidly. In photonic crystals, distributed scattering defines the band structure and defect-mode response. In plasmonic crystals, lattice resonances and hybridized particle modes cannot be assigned to isolated particles alone. In van der Waals heterostructures, polaritonic dispersion can reshape optical modes at room temperature. In coupled-resonator systems, the target mode may exist only because of mutual field interaction across the structure. Nonlocality is therefore both a limitation and a resource. It limits the direct transfer of local or short-range design rules to realistic optical devices, but it also supplies additional control variables for engineering collective modes in topological structures, metasurfaces, photonic crystals, plasmonic crystals, van der Waals heterostructures and hybrid resonator platforms. A useful nanophotonic design should therefore answer four questions: what interaction model is the geometry actually realizing, what finite electromagnetic observable supports the claimed function, how stable is that observable under nonlocal coupling, loss, detuning and fabrication uncertainty, and where does the design lie in the confidence landscape between topological and trivial response? Only then can topology, inverse design and physics-informed learning move from ideal labels toward calibrated statements about real electromagnetic devices.